\documentclass[12pt]{revtex4}
\usepackage{dcolumn}
\usepackage{graphicx}
\usepackage{amsmath}
\usepackage{amsfonts}
\usepackage{amssymb}
\usepackage{psfrag}
\usepackage{wrapfig}
\usepackage{subfigure}
\usepackage{makeidx}
\usepackage{bm}
\usepackage{epsf}
\usepackage{hyperref}

\begin{document}

\title{Thermodynamics of Slowly Rotating Charged Black Holes in anti-de Sitter
Einstein-Gauss-Bonnet Gravity}

\author{ De-Cheng Zou$^{1}$, Zhan-Ying Yang$^{1}$\footnote{Email:zyyang@nwu.edu.cn}
and Rui-Hong Yue$^{2}$\footnote{Email:yueruihong@nbu.edu.cn}}
\affiliation{ $^{1}$Department of Physics, Northwest University, Xi'an, 710069, China\\
$^{2}$Faculty of Science, Ningbo University, Ningbo 315211, China}

\date{\today}

\begin{abstract}
\indent

By using a new approach, we demonstrate analytic expressions for slowly rotating
Gauss-Bonnet charged black holes with negative cosmological constant.
Up to the linear order of the rotating parameter $a$,
the mass, Hawking temperature and entropy of the charged black holes get
no corrections from rotation.
\end{abstract}



\maketitle

\section{Introduction}
\indent

The thermodynamics of black holes in anti-de Sitter spaces have attracted a great deal of
attention. one reason for this is the role of AdS/CFT. It is well-known that
the AdS Schwarzschild black hole is thermodynamically unstable when the horizon radius is
small, while it is stable for large radius; there is a phase transition, named Hawking-Page
transition \cite{Hawking:1982dh}, between the large stable black hole and a thermal AdS
space. This phase transition is explained by Witten as the confinement/deconfinement
transition of Yang-Mills theory in the AdS-CFT correspondence \cite{Witten:1998zw}.
Since the thermodynamics of black holes has a deep connect with quantum theory of gravity,
they may be modified by using other theories of gravity with higher derivative curvature
terms. In the AdS/CFT correspondence, the higher derivative terms can be regarded as
the corrections of large N expansion in the dual conformal field theory. In general,
the higher powers of curvature can give rise to a fourth or even
higher order differential equation for the metric, and it will introduce ghosts and
violate unitarity. So, the higher derivative terms may be a source of inconsistencies.
However, Zwiebach and Zumino \cite{Zwiebach:1985uq}
found that the ghosts can be avoided if the higher derivative terms only consist of the
dimensional continuations of the Euler densities, leading to second order field
equations for the metric. This higher derivative theory is so-called Lovelock
gravity \cite{Lovelock:1971yv}, and the equations of motion contain the most symmetric
conserved tensor with no more than second derivative of the metric. In this paper, we
restrict ourselves to explore the first three terms of the Lovelock gravity,
corresponding to the cosmological constant, Einstein and Gauss-Bonnet terms respectively.
In recent years, there have been considerable works for understanding the role of
the higher curvature terms from various points of view, especially for higher
dimensional black hole physics. the analytic expression of static and spherically
symmetric Gauss-Bonnet black hole solutions have been investigated in \cite{Boulware:1985wk}.
The thermodynamics of the uncharged static spherically Gauss-Bonnet black hole
solutions have been considered in \cite{Cai:2003gr} and of charged
solutions in \cite{Myers:1988ze, Cvetic:2001bk}.

It is of interest to generalize these static and spherically symmetric black hole
solutions by including the effects of rotation. In the AdS/CFT correspondence,
the rotating black holes in AdS space are dual to certain CFTs in a rotating
space \cite{Hawking:1998kw}, while charged ones are dual to CFTs with chemical
potential \cite{Cvetic:1999rb}. In general relativity, the most general higher
dimensional rotating black holes in AdS space have been investigated
in \cite{Gibbons:2004uw}. While,
since the equations of motion of Lovelock gravity are highly nonlinear,
it is rather difficult to obtain the explicit rotating black hole solutions.
Recently, some numerical results about the existence of five-dimensional rotating
Gauss-Bonnet black holes with angular momenta of the same magnitude have been presented
in \cite{Brihaye:2008kh}. Besides, it is worth to mention that some rotating black
brane solutions have been investigated in Gauss-Bonnet gravity \cite{Dehghani2006}.
Nevertheless, these solutions are essentially obtained by a Lorentz boost from
corresponding static ones. They are equivalent to static ones locally,
although not equivalent globally.
In order to find rotating black hole solutions in the presence
of dilaton coupling electromagnetic field in Einstein(-Maxwell) theory, Horne and
Horowitz \cite{Horne:1992zy} first developed a simple
method that a small angular momentum as a perturbation was introduced into a
non-rotating system, and obtained slowly rotating dilaton black hole solutions.
Such so-called slowly rotating black holes have been extensively discussed
in general relativity \cite{Sheykhi:2008rm}.
With the help of it, Kim and Cai \cite{Kim:2007iw} presented analytic solutions of
slowly rotating uncharged Gauss-Bonnet black holes with one non-vanishing
angular momentum and arrived at the asymptotic forms of $g(r)$ and $c(r)$ in
charged case, here the rotating parameter $a$ appears as a small quantity.
In this paper, we will also analyze slowly rotating charged Gauss-Bonnet black holes.
We find that the off diagonal component of the stress-tensor of electromagnetic
field is not independent of $c(r)$ in charged case. Then, the equations
for $g(r)$ and $c(r)$ become two non-homogeneous differential equations.
But, it's possible to get analytic expressions for $g(r)$ and $c(r)$.

The outline of this paper is as follows. In section \ref{2s}, we review Gauss-Bonnet
gravity, and obtain the equations of gravitation and electromagnetic fields.
By putting a new metric into these equations, the slowly rotating charged black hole
solution $f(r)$ and exact expressions for functions $g(r)$ and $c(r)$ are obtained.
Later, some related physical properties of black holes are studied there.
We finish this paper with some concluding remarks.

\section{Slowly Rotating Black Holes in AdS Space\label{2s}}
\subsection{Action and Black Hole Solutions}
\indent

The action of Gauss-Bonnet gravity in the presence of electromagnetic field can be
written as
\begin{eqnarray}
{\cal I}=\frac{1}{16\pi G}\int d^{D}x\sqrt{-g}(-2\Lambda+{\cal L}_{1}+\alpha {\cal L}_{2}
-4\pi G F_{\mu\nu}F^{\mu\nu}) ,\label{1a}
\end{eqnarray}
where $\alpha$ is the Gauss-Bonnet coefficient with dimension $(length)^2$,
$F_{\mu\nu}=\partial_{\mu}A_{\nu} -\partial_{\nu}A_{\mu}$ is
electromagnetic field tensor with a vector potential $A_{\mu}$. The Einstein term ${\cal L}_1$
equals to $R$, and the Gauss-Bonnet term ${\cal L}_2$ is
$R_{\mu\nu\sigma\kappa}R^{\mu\nu\sigma\kappa}-4R_{\mu\nu}R^{\mu\nu}+R^2$.
It is easy to find that the solution is asymptotically flat for $\Lambda=0$, AdS for
negative value of $\Lambda$ and dS for positive value
of $\Lambda$. We discuss the case of asymptotically AdS solutions in this paper.

Varying the action with respect to the metric tensor $g_{\mu\nu}$ and electromagnetic tensor
field $F_{\mu\nu}$, the equations for gravitation and electromagnetic fields are
\begin{eqnarray}
\Lambda g_{\mu\nu}+G^{(1)}_{\mu\nu}+\alpha G^{(2)}_{\mu\nu}=8\pi GT_{\mu\nu}, \label{2a}
\end{eqnarray}
\begin{eqnarray}
\partial_{\mu}(\sqrt{-g}F^{\mu\nu})=0.\label{3a}
\end{eqnarray}
Here $T_{\mu\nu}=F_{\mu\alpha}F_{\nu}^{~\alpha}
-\frac{1}{4}g_{\mu\nu}F_{\alpha\beta}F^{\alpha\beta}$
is the energy-momentum tensor of electromagnetic field, $G^{(1)}_{\mu\nu}=R_{\mu\nu}
-\frac{1}{2}Rg_{\mu\nu}$ is Einstein tensor, and $G^{(2)}_{\mu\nu}$ is
Gauss-Bonnet tensor given as
\begin{eqnarray}
G^{(2)}_{\mu\nu}=2(R_{\mu\sigma\kappa\tau}R_{\nu}^{~\sigma\kappa\tau}
-2R_{\mu\rho\nu\sigma}R^{\rho\sigma}-2R_{\mu\sigma}R^{\sigma}_{~\nu}
+RR_{\mu\nu})-\frac{1}{2}{\cal L}_{2}g_{\mu\nu}.\label{4a}
\end{eqnarray}

Usually, the action Eq.~(\ref{1a}) is supplemented with surface terms (a Gibbons-Hawking
surface term) whose variation will  cancel the extra normal derivative term in deriving
the equation of motion Eq.~(\ref{2a}). However, these surface terms is not necessary in
our discussion and will be neglected. For Gauss-Bonnet gravity, the nontrivial third term
requires the dimension(D) of spacetime satisfying $D \geq 5$.

We assume that the metric of slowly rotating spacetime is
\begin{eqnarray}
ds^2=-f(r)dt^2+\frac{1}{f(r)}dr^2+\sum^{D}_{i=j=3}r^2h_{ij}dx^idx^j
-2ar^2g(r)h_{44}dtd\phi,\label{5a}
\end{eqnarray}
where $h_{ij}dx^idx^j$ represents the metric of a $(D-2)$ dimensional hyper-surface
with constant curvature scalar $(D-2)(D-3)k$ and volume $\Sigma_{k}$, here k is a constant.
The latter is the unit metric on $S^{n-1}$, $R^{n-1}$, or $H^{n-1}$, respectively,
for $k=1$, 0 or -1.

For the convenience future, we introduce
\begin{eqnarray}
\Lambda=-\frac{(D-1)(D-2)}{2l^2},\quad \tilde{\alpha}=\alpha (D-3)(D-4).\label{6a}
\end{eqnarray}
For $g(r)=0$, we can arrive at the function $f(r)$ in charged case \cite{Cvetic:2001bk, Wiltshire:1985us}
\begin{eqnarray}
f(r)=k+\frac{r^2}{2\tilde{\alpha}}\Big[1-\sqrt{1-\frac{4\tilde{\alpha}}{l^2}
+\frac{4\tilde{\alpha}m}{r^{D-1}}-\frac{4\tilde{\alpha}q^2}{r^{2D-4}}}\Big],\label{7a}
\end{eqnarray}
where $m$ and $q$ are related to the total mass $M=\frac{(D-2)\Sigma_k}{16\pi G}m$
and total charge $Q^2=\frac{2\pi(D-2)(D-3)}{G}q^2$ of spacetimes.

In \cite{Kim:2007iw}, the slowly rotating Gauss-Bonnet black holes have been studied
in charged case. Kim and Cai reported expressions for $g(r)$ and $c(r)$ in
asymptotic form. However, it is possible to present the analytic expressions
for $g(r)$ and $c(r)$ by adopted a new approach.
In case of $g(r) \neq 0$, the metric function $f(r)$ still keep the form Eq.~(\ref{16a}).
Here, we introduce $f(r)=k-r^2\varphi_*$ with
\begin{eqnarray}
\varphi_*=-\frac{1}{2\tilde{\alpha}}\Big[1-\sqrt{1-\frac{4\tilde{\alpha}}{l^2}
+\frac{4\tilde{\alpha}m}{r^{D-1}}-\frac{4\tilde{\alpha}q^2}{r^{2D-4}}}\Big].\label{8a}
\end{eqnarray}

Since the black hole rotates along the direction $\phi$, it will generate a
magnetic field. Considering this effect, we get the gauge potential
\begin{eqnarray}
A_{\mu}dx^{\mu}=A_{t}dt+A_{\phi}d\phi.\label{9a}
\end{eqnarray}
Here we assume $A_{\phi}=-aQc(r)h_{44}$. As a result, the electro-magnetic field
associated with the solution are
\begin{eqnarray}
F_{tr}=-A_t',\quad F_{r\phi}=-a Q c'(r)h_{44},\quad
F_{\theta\phi}=-a Q c(r)h'_{44}.\label{10a}
\end{eqnarray}
where $Q$, an integration constant, is the electric charge of the black hole and a prime
denotes the derivative with respect to $r$. Form $t$-component of electromagnetic field
equation $\partial_{\mu}(\sqrt{-g}F^{\mu\nu})=0$, one can find $F_{tr}=\frac{Q}{4\pi r^{D-2}}$,
which is the same as the static form. Unlike the static case, there exist the $\phi$-component
of the electromagnetic field equation, and then the equation for function $c(r)$ reads \cite{Kim:2007iw}
\begin{eqnarray}
(r^{D-4}f(r)c'(r))'-2k(D-3)r^{D-6}c(r)=\frac{g'(r)}{4\pi}.\label{11a}
\end{eqnarray}

Meanwhile, there exist off diagonal $t\phi$ component of equations of motion. It was considered
that this equation decoupled from function $c(r)$ in \cite{Kim:2007iw}. While, one can easily
verify that $t\phi$ component is concerned with functions $g(r)$ and $c(r)$. A tedious computation
leads to a following equation
\begin{eqnarray}
r^D(1+2\tilde{\alpha}\varphi_*)g(r)'=4GQ^2c(r)+C_3.\label{12a}
\end{eqnarray}

We substitute metric Eq.~(\ref{5a}) into the action Eq.~(\ref{1a}).
It is interesting to notice that the action Eq.~(\ref{1a}) is in the absence of parameter $a$
and $\varphi_*$ is determined by solving for the real roots of the following second-order
polynomial equation \cite{Myers:1988ze}
\begin{eqnarray}
\frac{1}{l^2}+\varphi_*+\tilde{\alpha}\varphi_*^2=\frac{m}{r^{D-1}}-\frac{q^2}{r^{2D-4}}.\label{13a}
\end{eqnarray}
We can easily verify by drawing a parallel between the Eqs.(\ref{12a}) and (\ref{13a})
\begin{eqnarray}
[m(1-D)+\frac{2q^2(D-2)}{r^{D-3}}]\frac{g(r)'}{\varphi_*'}=4GQ^2c(r)+C_3.\label{14a}
\end{eqnarray}
Let the constant $C_3=m(D-1)$ and $g(r)=-\varphi_*$, one can find two explicit solutions
for functions $g(r)$ and $c(r)$
\begin{eqnarray}
g(r)&=&-\varphi_*\nonumber\\
&=&\frac{1}{2\tilde{\alpha}}\Big[1-\sqrt{1-\frac{4\tilde{\alpha}}{l^2}
+\frac{4\tilde{\alpha}m}{r^{D-1}}-\frac{4\tilde{\alpha}q^2}{r^{2D-4}}}\Big]\label{15a}\\
c(r)&=&-\frac{1}{4\pi(D-3)r^{D-3}}.\label{16a}
\end{eqnarray}
Apparently the expressions for $g(r)$ and $c(r)$ still satisfy
the $\phi$-component of the electromagnetic field equation Eq.~(\ref{11a}).

\subsection{Physical Properties}
\indent

In this subsection, we analyze the physical properties of slowly rotating charged black holes.
Shown in Eq.~(\ref{16a}), the slowly rotating charged solution $f(r)$ is independent of $a$.
Most interesting physical properties depend only on $a^2$, but one can
still extract some useful information from it.

For slowly rotating solution, the horizon $r_+$ is still determined by the equation $f(r_+)=0$.
Then, $r_+$ is related to the mass $(m)$ and charge $(q)$ of the black hole by the relation
\begin{eqnarray}
r_+^{2D}(r_{+}^4/l^2+kr_{+}^2+k^2\tilde{\alpha})=mr_+^{D+5}-q^2r_+^8.\label{17a}
\end{eqnarray}
So, one can write the gravitational mass of black holes
\begin{eqnarray}
M=\frac{(D-2)\Sigma_{k}r_{+}^{D-1}}{16\pi G}[\frac{1}{l^2}+kr_{+}^{-2}
+k^2\tilde{\alpha}r_+^{-4}+\frac{q^2}{r_+^{2D-4}}].\nonumber\\\label{18a}
\end{eqnarray}
Then, the Hawking temperature of the black hole can be obtained by required the absence of
conical singular at the horizon in the Euclidean of the black hole solution. It is the same as
the static case
\begin{eqnarray}
T=\frac{f'(r_{+})}{4\pi}=\frac{\Gamma(r_+)}{4\pi r_{+}(r_{+}^2+2k\tilde{\alpha})},\label{19a}
\end{eqnarray}
where $\Gamma(r_+)=(D-1)r_{+}^4/l^2+(D-3)kr_{+}^2+(D-5)k^2\tilde{\alpha}-(D-3)q^2/r_+^{2D-8}$.
Thus, the angular momentum of the black hole
\begin{eqnarray}
J=\frac{2aM}{D-2}=\frac{a\Sigma_{k}r_{+}^{D-1}}{8\pi G}[\frac{1}{l^2}+kr_{+}^{-2}
+k^2\tilde{\alpha}r_+^{-4}+\frac{q^2}{r_+^{2D-4}}].\label{20a}
\end{eqnarray}

Usually, the entropy of black hole satisfies the so-called area law of entropy which states
that the black hole entropy equals to one-quarter of the horizon
area \cite{Gibbons:1977mu}, \cite{Hawking:1974rv}. It applies to all kinds of black
holes and black strings of Einstein gravity \cite{Hunter:1998qe}.
However, in higher derivative gravity, the area law of the entropy is not satisfied
in general \cite{Jacobson:1993xs}. Since black hole can be regard as a thermodynamic system,
it obeys the first law of thermodynamics $dM=TdS+\omega_{H}dJ$.
Through the angular velocity $\omega_{H}$, one can get the entropy of black hole.

For the slowly rotating solution, the stationarity and  rotational symmetry
metric Eq.~(\ref{5a}) admits two commuting Killing vector fields
\begin{eqnarray}
\xi_{(t)}=\frac{\partial}{\partial t}, \quad \xi_{\phi}=\frac{\partial}{\partial \phi}\label{21a}.
\end{eqnarray}
The various scalar products of these Killing vectors can be expressed through the metric
components as follows
\begin{eqnarray}
\xi_{(t)}\cdot\xi_{(t)}&=&g_{tt}=-f(r),\nonumber\\
\xi_{(t)}\cdot\xi_{(\phi)}&=&g_{t\phi}=-ar^2g(r)h_{44},\nonumber\\
\xi_{(\phi)}\cdot\xi_{(\phi)}&=&g_{\phi\phi}=r^2h_{44}.\nonumber
\end{eqnarray}

To examine further properties of the slowly rotating black holes, as well as physical
processes near such a black hole, we introduce a family of locally non-rotating observers.
The coordinate angular velocity for these observers that move on orbits with constant $r$
and $\theta$ and with a four-velocity $u^{\mu}$ such that $u\cdot\xi_(\phi)=0$
is given by \cite{Aliev:2007qi},
\begin{eqnarray}
\Omega&=&-\frac{g_{t\phi}}{g_{\phi\phi}}=ag(r)\nonumber\\
&=&\frac{a}{2\tilde{\alpha}}\Big[1-\sqrt{1-\frac{4\tilde{\alpha}}{l^2}
+\frac{4\tilde{\alpha}m}{r^{D-1}}-\frac{4\tilde{\alpha}q^2}{r^{2D-4}}}\Big].\label{22a}
\end{eqnarray}
In contrast to the case of an ordinary kerr black hole in asymptotically flat spacetime,
the angular velocity does not vanish at spatial infinity
\begin{eqnarray}
\Omega_\infty=\frac{a}{2\tilde{\alpha}}\Big(1-\sqrt{1-\frac{4\tilde{\alpha}}{l^2}}\Big)
=\frac{a}{l_{eff}^2}.\label{23a}
\end{eqnarray}

When approaching the black hole horizon, the angular velocity turns to be
$\Omega_H=ag(r_+)=-a\varphi(r_+)=-\frac{ak}{r_+^2}$. This $\Omega_H$ can be thought as
the angular velocity of the black hole. The relative angular velocity  with respect to
a frame static at infinity is defined by
\begin{eqnarray}
\omega_H=\Omega_H-\Omega_\infty=-a(\frac{k}{r_+^2}+\frac{1}{l_{eff}^2}).\label{24a}
\end{eqnarray}
Therefore, we get the entropy of slowly rotating black hole up to the linear order of
the rotating parameter $a$
\begin{eqnarray}
S=\frac{\Sigma_{k}}{4G}r_{+}^{D-2}[1+\frac{2(D-2)k\tilde{\alpha}}{(D-4)r_{+}^2}].\label{25a}
\end{eqnarray}

\section{Concluding Remarks}
\indent

By introducing a small angular momentum, we discussed slowly rotating Gauss-Bonnet
charged black holes. For charged case, the vector potential has an extra nonradial
component $A_{\phi}=-aQc(r)h_{44}$ due to the rotation of the black holes.
In addition, since the off-diagonal component of the stress-tensor of electro-magnetic
field was related to $c(r)$, the equations for $g(r)$ and $c(r)$ become two
non-homogeneous differential equations. Then, the analytic solutions for $c(r)$
and $g(r)$ have been separately expressed as $c(r)=-\frac{1}{4\pi(D-3)r^{D-3}}$
and $g(r)=-\varphi_*$ by using a new approach, while the function $f(r)$
still keep the form of the static solution. Up to the linear order of the rotating
parameter $a$, the expressions of the mass, temperature, and entropy for the
charged black holes got no correction from rotation.

{\bf Acknowledgment }

This work has been supported by the National Natural Science Foundation
of China under grant No. 10875060 and 10975180.

\end{document}